\documentclass{article}

\usepackage{arxiv}

\usepackage[utf8]{inputenc} 
\usepackage[T1]{fontenc}    
\usepackage{hyperref}       
\usepackage{url}            
\usepackage{booktabs}       
\usepackage{amsfonts}
\usepackage{amssymb,dsfont}
\usepackage{amsmath}       
\usepackage{nicefrac}       
\usepackage{microtype}      
\usepackage{lipsum}		
\usepackage{graphicx}
\usepackage{doi}
\usepackage{cite}

\title{Optical MIMO Communication With Unequal Power Allocation To Channels}

\author{ \href{https://orcid.org/0000-0001-7000-4531}{\includegraphics[scale=0.06]{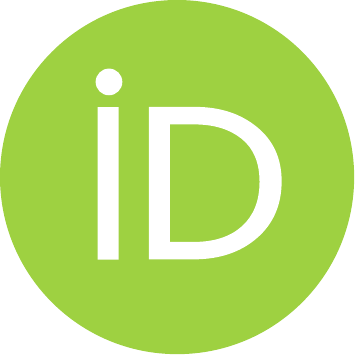}\hspace{1mm}Aritra Laha} \hspace{1cm}
	\href{https://orcid.org/0000-0002-8405-7308}{\includegraphics[scale=0.06]{orcid.pdf}\hspace{1mm}Santosh Kumar} \thanks{Aritra Laha and Santosh Kumar are with Shiv Nadar Institution of Eminence, Uttar Pradesh, India. }
}

\date{}

\renewcommand{\headeright}{}
\renewcommand{\undertitle}{}

\hypersetup{
pdftitle={A template for the arxiv style},
pdfsubject={q-bio.NC, q-bio.QM},
pdfauthor={David S.~Hippocampus, Elias D.~Striatum},
pdfkeywords={First keyword, Second keyword, More},
}

\begin{document}
\maketitle

\begin{abstract}
	 Multiple input multiple output (MIMO) approach in fiber optical communication has emerged as an effective proposition to address the ever increasing demand for information exchange. In the ergodic case, the multiple channels, associated with multiple modes or cores or both in the optical fiber, is modeled by the Jacobi ensemble of random matrices. A key quantity for assessing the performance of MIMO systems is the mutual information (MI). We focus here on the case of an arbitrary transmission covariance matrix and derive exact determinant based results for the moment generating function (MGF) of mutual information (MI), and thereby address the scenario of unequal power per excited mode. The MGF is used to obtain Gaussian- and Weibull-distribution based approximations for the probability density function (PDF), cumulative distribution function (CDF) or, equivalently, the outage probability, and also the survival function (SF) or reliability function. Moreover, a numerical Fourier inversion approach is implemented to obtain the PDF, CDF, and SF directly from the MGF. The MGF is further used to investigate the ergodic capacity, which is the first moment (mean) of the mutual information. The analytical results are found to be in excellent agreement with Monte Carlo simulations. Our study goes beyond the earlier investigations where covariance matrix proportional to identity matrix has been considered which corresponds to equal power allocation per excited mode.
\end{abstract}


\section{Introduction}\label{sec1}
Optical fiber communication has played a significant role in revolutionizing the telecommunications industry~\cite{Keiser2000,Agarwal2002,FS2005,Reinhold2016,SSB2008}. It has enabled telecommunications links to be established over much greater distances and at the same time maintaining lower levels of loss in the transmission medium. Furthermore, in general, compared to wireless communication and other communication means, it supports a much higher data transmission rate. However, due to the steady growth in demand for bandwidth, the traditional view of optical fibers being able to provide unlimited bandwidth seems to no longer hold. Challenges like dispersion, noise, and nonlinearities have to be efficiently dealt with. Therefore, the focus has shifted to develop more sophisticated signal processing techniques which can be implemented in optical fiber communication and thereby lead to the advancement of low-cost, high-data-rate optical communication systems. To this end, along with other techniques, much research has been devoted into optimizing multiplexing in time, wavelength, polarization, and phase. The corresponding techniques are referred to as time-division multiplexing (TDM), wavelength-division multiplexing (WDM), polarization-division multiplexing (PDM), and phase modulation (PM)~\cite{Keiser2000,Agarwal2002,FS2005,Reinhold2016,SSB2008,RFNL2013}. Yet another approach to enhance the capacity of optical communication is via space division multiplexing (SDM), which relies on tightly packing several spatial channels with the aid of multimode or multicore fibers~\cite{RFNL2013,WWP2019}. In this context, one has to naturally deal with the multiple-input multiple-output (MIMO) approach in optical communication. It offers several advantages such as higher data rate, lower bit error rate (BER), higher reliability, lower susceptibility of tapping, and supports a large number of users. While it does come with some drawbacks also, such as the requirement of complicated hardware and software, etc., the advantages weigh more and consequently MIMO approach has attracted a lot of interest in the recent years~\cite{KSK2009,WF2011,ARWRP2012,OKNM2012,DFS2012,DFS2013,KMV2014,WWP2019,NB2017,ND2020,WZG2018,WLLB2019}.

The MIMO implementation in wireless communication has been exhaustivelystudied~\cite{FG1998,Telatar1999,Muller2002,CWZ2003,MC2005,MGC2007,KP2010,TV2004,CD2011,MA2007,SMM2006,AIK2013,AKW2013,PKFD2017,ACNZ2018} since the seminal works of Foschini and Gans~\cite{FG1998}, and Telatar~\cite{Telatar1999}. However, the optical MIMO communication, especially from a statistical viewpoint, has remained comparatively an unexplored territory. Random matrix theory~\cite{Mehta2004,Forrester2010} provides a powerful statistical approach for modeling the channel matrix in both wireless and fiber-optics MIMO channel communications~\cite{TV2004,CD2011,DFS2012,DFS2013}. The wireless communication inherently has the signal fading associated with it due to propagation of the signal in the atmosphere or free space. Accordingly, the channel matrix describing the fading is modeled using the Wishart-Laguerre random matrix~\cite{Mehta2004,Forrester2010}. Moreover, other generalized scenarios of wireless communication, such as progressive scattering, multiple users, etc. are modeled by random matrix ensembles which involve various compositions of Wishart-Laguerre matrices, such as their products and sums~\cite{AIK2013,AKW2013,KP2010,PKFD2017}. In contrast, owing to the one-dimensional, near-lossless propagation through the optical fiber, the signal propagation through it can be analyzed using the unitary scattering matrix approach. In the case of optical fiber communication, the multiple channels correspond to multiple modes or multiple cores or both in the fiber~\cite{KSK2009,WF2011,ARWRP2012,OKNM2012,DFS2012,DFS2013,KMV2014,WWP2019}.  It has been shown in~\cite{WF2011,DFS2012,DFS2013,KMV2014} that the channel matrix in such a scenario is a truncation of the unitary transmission matrix and is described by the classical Jacobi ensemble of random matrices~\cite{Mehta2004,Forrester2010} in the ergodic case.

One of the key quantities to assess the performance of MIMO systems is the mutual information (MI)~\cite{Telatar1999}. For the optical MIMO communication, the exact first moment (mean) of MI has been calculated by Dar {\it et al.} in~\cite{DFS2012,DFS2013}. A closed-form expression for the outage probability has been derived by Karadimitrakis {\it et al.} in~\cite{KMV2014}, which is convenient to use when small dimension matrices are involved. Additionally, therein, large-dimension asymptotic results for the mean and variance of MI, and the outage probability have been provided. Higher order moments of MI in the high signal-to-noise ratio (SNR) regime have been calculated by Wei {\it et al.} in~\cite{WZG2018,WLLB2019}. Upper and lower bounds for the ergodic capacity of MIMO Jacobi fading channels have been obtained by Nafkha and Bonnefoi in~\cite{NB2017}. Moreover, very recently, Nafkha and Demni have provided closed-form expressions for the ergodic capacity and MMSE achievable sum rate~\cite{ND2020}. All these studies have considered an equal power allocation for the excited modes used for transmission. In the present work, we focus on the case of an arbitrary positive-semidefinite transmission covariance matrix, and hence address the scenario of an unequal power allocation. We derive an exact determinantal expression for the moment generating function (MGF) of MI, which can be used to evaluate the corresponding moments. Based on the moment or cumulant matching approach, the MGF is used to obtain Gaussian- and Weibull-distribution based approximations for the probability density function (PDF), cumulative distribution function (CDF), and survival function (SF). It should be noted that the CDF is equivalent to the outage probability. Additionally, direct numerical evaluations of PDF, CDF, and SF from the MGF are also implemented using a numerical Fourier-inversion approach. 

The presentation scheme in the rest of the paper is as follows. In Section~\ref{sec2}, we derive exact closed form expressions for the MGF of mutual information, which also gives access to the corresponding moments and cumulants. In Section~\ref{sec3}, we obtain Gaussian- and Weibull-distribution based approximations for the PDF, CDF, and SF of the mutual information. Moreover, a very simple yet effective numerical Fourier-inversion approach is adopted to obtain the distribution of the mutual information. Section~\ref{sec4} is devoted to presenting evaluations of the results and their interpretation. We conclude with a brief summary of our work in Section~\ref{sec5}.

\section{Calculation of MGF}\label{sec2}
The MIMO channel is described by the equation~\cite{Telatar1999},
\begin{equation} 
\bf {y} = \bf{Hx} +\bf{z},
\end{equation}
 where the channel matrix \textbf{H} is an ($ n\times m $)-dimensional matrix, and \textbf{x} and \textbf{y} are the $m$-dimensional input and $n$-dimensional output signal vectors, respectively. Also, \textbf{z} is the $n$-dimensional zero-mean unit-variance complex Gaussian noise vector. In the case of optical MIMO communication, the channel matrix \textbf{H} happens to be a submatrix of an ($ l \times l $)-dimensional  unitary transmission matrix, where $l$ is the total number of available channels in the optical fiber~\cite{DFS2012,DFS2013,KMV2014}. The parameters $m$ and $n$ are decided by the number of channels excited at the input and output sides of the fiber, respectively. Our aim is to explore the statistics of MI, which is given by~\cite{Telatar1999},
\begin{equation}
\label{mutualinfo}
I = \ln \det(\mathds{1}_n + {\bf{HQH}}^\dag) =\ln \det(\mathds{1}_m + \bf{QH^\dag H}),
\end{equation}
where the last term follows using the Sylvester's identity~\cite{Sylvester1851}. The ($m\times m$)-dimensional transmission covariance matrix, {\bf Q} = $\mathbb{E}[{\bf xx}^\dag]$, takes care of the transmission power constraint~\cite{DFS2013}. So far, investigations pertaining to the optical MIMO channel have focussed on the case of matrix \textbf{Q} proportional to the identity matrix $\mathds{1}_m$. In this work, we consider it to be an arbitrary positive-semidefinite matrix. It has been shown in~\cite{DFS2013}, that the results for $l< m+n$ case can be obtained from those of $l\ge m+n$. Therefore, we focus on the latter in the following.  It is known that the distribution of $\bf{H^\dag H}$ coincides with that of the Jacobi ensemble of random matrices~\cite{Mehta2004, Forrester2010},
\begin{align}
\label{Hdist}
\mathcal{P}({\bf{H^\dag H}})\propto\det({\bf{H^\dag H}})^{|m-n|} \det(\mathds{1}_m - {\bf{H^\dag H}})^{l-m-n},
\end{align}
and hence the name \emph{Jacobi-MIMO} is used in the context of optical communication~\cite{DFS2013}. It might be worth mentioning at this point that the Jacobi ensemble of random matrices has remained a prominent model for describing electronic transport properties in chaotic mesoscopic cavities~\cite{BM1994,JPB1994,Beenakker1997,Forrester2006,SM2006,KP2010a,KP2010b,FK2019}. However, its implementation for modeling MIMO channel in fiber optics communication is relatively recent~\cite{DFS2012,DFS2013,KMV2014}.

For $m\le n$, the joint probability density function (PDF) of the unordered eigenvalues $(0<\lambda_i<1)$ of ${\bf{H^\dag H}}$ is given by~\cite{KMV2014,DFS2012,DFS2013},
\begin{equation}
\label{P1L}
p(\{\lambda\}) = C_{m,n}~\Delta_{m}^{2}(\{\lambda\}) \prod_{i=1}^{m} \lambda_i^{n-m} (1 - \lambda_i)^{l-m-n}.
\end{equation}
In the above equation,
\begin{equation} 
\Delta_{\alpha}(\{\lambda\}) :=\det[\lambda_k^{j-1}]_{j,k=1,..,\alpha}= \prod_{1 \leq j < k\leq \alpha} (\lambda_k - \lambda_j)
\end{equation}
 is the Vandermonde determinant, and the normalization factor is given by 
 \begin{equation}
 C_{\alpha,\beta} := \prod_{i=1}^{\alpha}\Gamma(l-i+1)/ [ \Gamma(i+1) \Gamma(l-\beta-i+1) \Gamma(\beta-i+1)],
 \end{equation}
 with $\Gamma(z)$ being the Gamma function~\cite{AS1972}.
For $m>n$, the rank $n$ random matrix ${\bf{H^\dag H}}$ has $n$ eigenvalues with joint probability density as in Eq.~\eqref{P1L} but with $m$ and $n$ interchanged. Additionally, it possesses $m-n$ eigenvalues identical to zero. The joint eigenvalue density in this case can be written with the help of Dirac delta function $\delta(\cdot)$ as,
\begin{equation}
\label{P2L}
p(\{\lambda\})=\frac{C_{n,m}}{m!} \Big[ \prod_{i=1}^{n} \lambda_{i}^{m-n} (1-\lambda_{i})^{l-m-n} \Delta_{n}^{2}(\{\lambda\})\prod_{j=n+1}^{m} \delta(\lambda_{j})+ m!\text{~permutations over } \{\lambda\}\Big].                    
\end{equation}
In the calculations involving this density below, the $m-n$ eigenvalues will be assigned the value zero in a limiting manner.

The MGF or characteristic function of the MI is given by
\begin{equation}
\label{MGFMI}
M(\kappa) = \mathbb{E}_I[\exp(\imath \kappa I)] =  \int_{-\infty}^{\infty} e^{\imath\kappa I}~ p(I)~ dI,
\end{equation}
where $\imath = \sqrt{-1}$ is the imaginary unit, and the averaging $\mathbb{E}_I$ is with respect to the PDF of $I$. The above definition is based on the Fourier transform of $p(I)$. Equivalently, we could consider the Laplace transform. However, we have found that numerical inversion to obtain the mutual information distribution from the MGF is numerically stable and much easier to implement for the Fourier transform than for the Laplace transform. Now, from the MGF, one can calculate the $\nu$-th order moment using the relation
\begin{equation}
\label{mmnt}
\mu_{\nu} = \frac{1}{\imath^{\nu}} \frac{d^{\nu}}{d\kappa^{\nu}} M(\kappa)|_{\kappa=0}.
\end{equation}
The ergodic capacity is the mean (first moment) of the mutual information and therefore corresponds to $\nu=1$. The second moment is obtained for $\nu=2$. The first cumulant is same as the mean, i.e., $\mu_1$, while the second cumulant is the variance $\sigma^2=\mu_2-\mu_1^2$.

\begin{figure}[!h]
\includegraphics[width=0.8\linewidth]{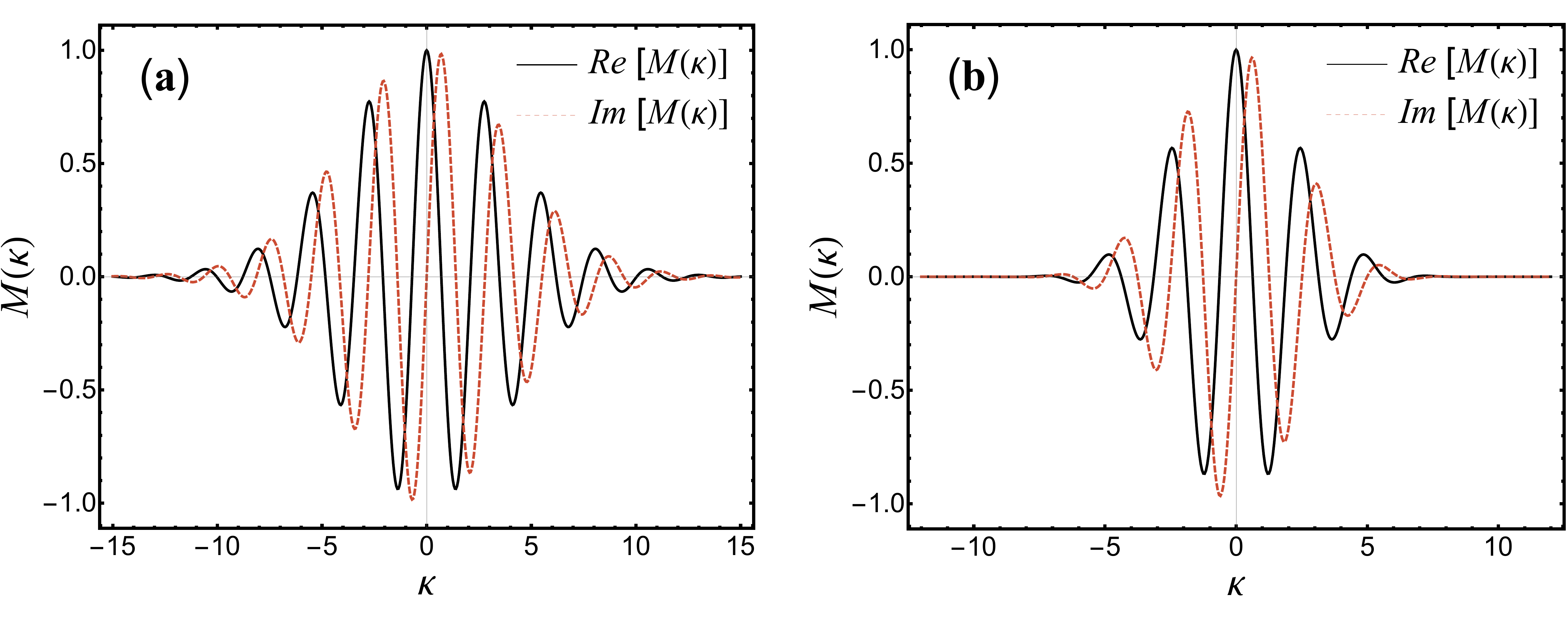}
\centering
\caption{\textit{Plots of real and imaginary parts of MGF of mutual information for (a) $m\le n$, with $(m,n,l)=(3,6,12)$, $(q_1,q_2,q_3)=(6.80, 1.50, 0.70)$, and (b) $m> n$, with $(m,n,l)=(4,3,10)$, $(q_1,q_2,q_3,q_4)=(11.00,5.00,1.50,0.50)$.}}
\label{mgftrunc}
\end{figure}

With the aid of Eqs.~\eqref{mutualinfo} and~\eqref{Hdist}, the MGF may also be evaluated using an averaging over the distribution of matrix ${\bf{H^\dag H}}$ as
\begin{align}
\label{mgfn}
M(\kappa)  =  \int \det(\mathds{1}_m + {\bf{QH^\dag H}})^{\imath \kappa} ~ \mathcal{P}({\bf{H^\dag H}})~ d[{\bf{H^\dag H}}].
\end{align}
We should remark that in the equal power allocation case, the $\bf{Q}$ matrix is proportional to the identity matrix. This makes the above expression for $M(\kappa) $ unitarily invariant and the calculation can be performed only using the eigenvalue statistics of $\bf{H^\dag H}$. However, in the case of unequal power allocation, the matrix $\bf{Q}$ no longer commutes with ${\bf H}$ and therefore the averaging must be performed with respect to both eigenvectors and eigenvalues, thereby making the problem nontrivial.

We use the eigenvalue decomposition of the Hermitian matrix $\bf{H^\dag H}$, viz. $\bf{H^\dag H}=\bf{U^\dag \Lambda U}$, where ${\bf U}$ is an $m$-dimensional unitary matrix and ${\bf \Lambda}$ is a diagonal matrix containing the eigenvalues of $\bf{H^\dag H}$. This enables us to transform the above matrix integral over $\bf{H^\dag H}$ into integrals over the eigenvalues $\{\lambda\}$ and the random unitary matrix ${\bf U}$. We have,
\begin{equation}
M(\kappa) = \int d\{\lambda\} \,p(\{\lambda\}) \int d{\bf U} \det(\mathds{1}_m + {\bf{QU^\dag \Lambda U}})^{\imath \kappa}.
\end{equation}
The integral over the random unitary matrix can be performed using the result given in~\cite{GR1989,Orlov2004} and leads us to
\begin{equation}
M(\kappa) = K_m(\kappa)\!\int d\{\lambda\} \,p(\{\lambda\}) \frac{\det[(1+q_j \lambda_k)^{\imath \kappa +m-1}]_{j,k=1}^m}{\Delta_m(\{q\}) \Delta_{m}(\{\lambda\})},
\end{equation}
where $\{q\}$ are the eigenvalues of ${\bf Q}$, and
\begin{equation}
K_m(\kappa) = (-1)^{m(m-1)/2}   ~~\prod_{i=1}^{m-1}\left(\frac{-\imath \kappa -m+i}{i}\right)^{i-m}.
\end{equation}
To proceed further, we need to examine how the number of transmitting channel ($m$) compares to that of the receiving channel ($n$).

\begin{figure}[!t]
\includegraphics[width=1\linewidth]{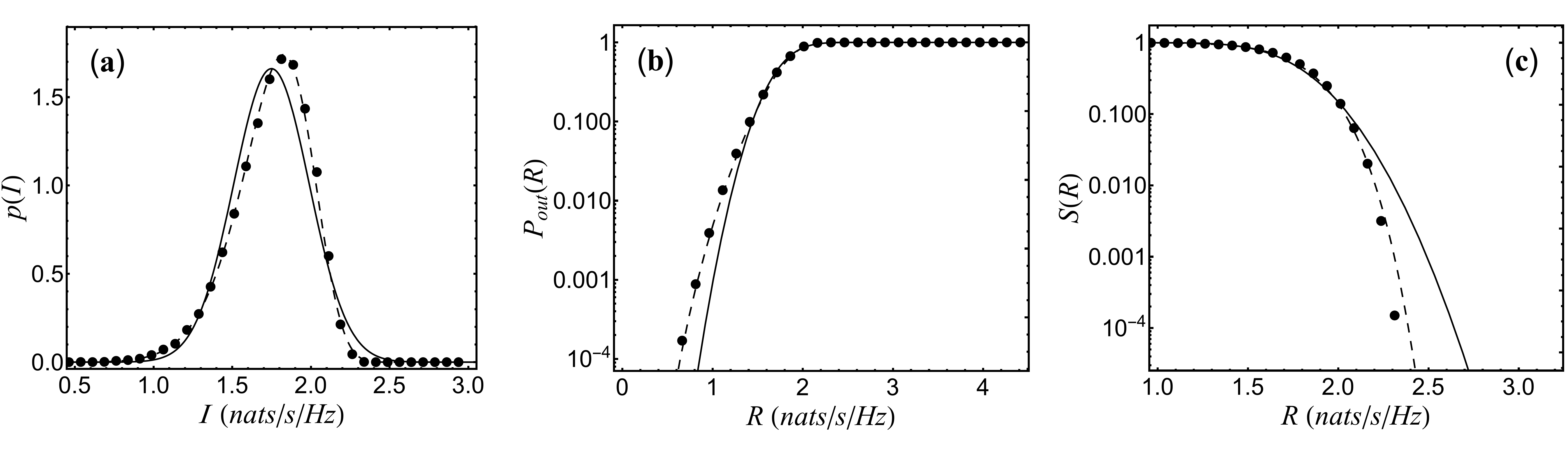}
\centering
\caption{\textit{Distribution of the mutual information for the case $m\le n$ with $(m,n,l)=(3,6,12)$ and $(q_1,q_2,q_3)=(8.80, 0.11, 0.09)$: (a) Probability density function (PDF), (b) outage probability or equivalently Cumulative distrbution function (CDF), and (c) survival function (SF). The solid line is based on the Gaussian approximation, the dashed line is using the Weibull approximation, and the disks are from Monte Carlo simulations.}}
\label{approx1}
\end{figure}
\begin{figure}[!t]
\includegraphics[width=1\linewidth]{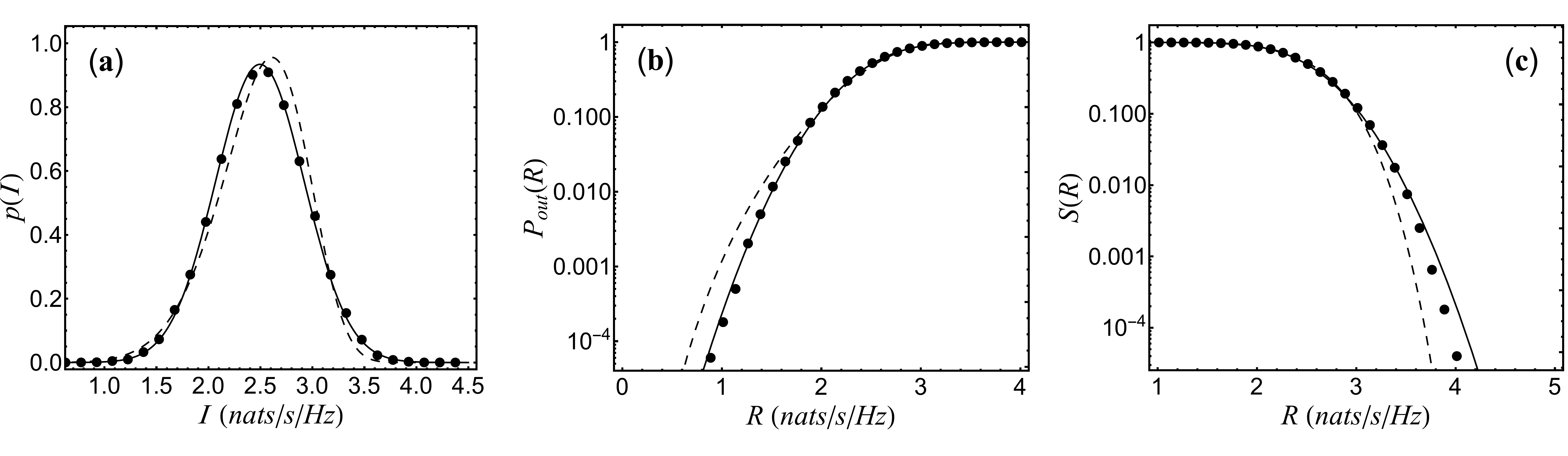}
\centering
\caption{\textit{Distribution of the mutual information for the case $m> n$ with $(m,n,l)=(4,3,10)$ and $(q_1,q_2,q_3,q_4)=(11.00,5.00,1.50,0.50)$: (a) PDF, (b) CDF, and (c) SF. The solid line is based on the Gaussian approximation, the dashed line is using the Weibull approximation, and the disks are from Monte Carlo simulations.}}
\label{approx2}
\end{figure}

\subsection{Case-I- $ m\le n $}\label{sec2.1}
In this case, we use Eq.~\eqref{P1L} and obtain
\begin{equation}
M(\kappa) = \frac{C_{m,n}K_m(\kappa)}{\Delta_m(\{q\})}\int d\{\lambda\} \prod_{i=1}^m\lambda_i^{n-m}(1-\lambda_i)^{l-m-n} \Delta_{m}(\{\lambda\})\det[(1+q_j \lambda_k)^{\imath \kappa +m-1}]_{j,k=1}^m,
\end{equation}
The integral over the eigenvalues can be performed using Andr\'{e}ief's formula~\cite{Andreief1883}, yielding us a closed-form expression of the MGF as a determinant,
\begin{equation}
\label{Mk1}
M(\kappa) = \frac{m! \,C_{m,n} K_m(\kappa)}{\Delta_m(\{q\})}\det[g_{j,k}(\kappa)]_{j,k=1}^m,
\end{equation}
where
\begin{align}
\nonumber
g_{j,k}(\kappa) &= \int_{0}^{1} d\lambda\,\lambda^{k+n-m-1}(1-\lambda)^{l-m-n}(1+q_j\lambda)^{\imath \kappa+m-1}  \\
            & = B(k+n-m,l-m-n+1)~ {}_2F_1(k+n-m,1-m-\imath \kappa;k+l+1-2m;-q_j).
\end{align}

Here, $B(z)$ is the Beta function and ${}_2F_1(a,b;c;z)$ is the Gauss hypergeometric function~\cite{AS1972}. Now, if some or all of the $q_j$ are equal, we will need to invoke a limiting procedure to obtain the corresponding analytical expressions. For numerical evaluation, however, Eq.~\eqref{Mk1} may be used by assigning the identical $q_j$'s very close but unequal numerical values. Moreover, the analytical result in the case of all $q$'s identical can be handled directly by noticing that ${\bf Q}=q\mathds{1}_{m}$. In this case, the eigenvalue and eigenvector dependences factorize and we do not require the nontrivial unitary group integral. The integral over the eigenvectors contributes to the overall constant only. Consequently, after the diagonalization, Eq.~\eqref{mgfn} can be expressed purely in terms of integration over the eigenvalues, i.e.,
\begin{equation}
M(\kappa) = \int d\{\lambda\} \,p(\{\lambda\}) \prod_{i=1}^{m}(1+q\lambda_i)^{\imath\kappa}.
\end{equation}
The joint PDF of eigenvalues is given by Eq.~\eqref{P1L} which can be used in the above expression and the integration can be performed with the aid of Andr\'{e}ief's formula~\cite{Andreief1883}. This gives the result for MGF in the case of equal power allocation $(q_1=\cdots=q_m=q)$ as, 
\begin{equation}
\label{Mk1eq}
M(\kappa) = m! \,C_{m,n} \det[t_{j,k}(\kappa)]_{j,k=1}^m,
\end{equation}
where,
\begin{align}
\nonumber
t_{j,k}(\kappa) &= \int_{0}^{1} d\lambda\,\lambda^{j+k+n-m-2}(1-\lambda)^{l-m-n}(1+q\lambda)^{\imath \kappa}  \\
            &= B(j+k+n-m-1,l-m-n+1)~ {}_2F_1(j+k+n-m-1,-\imath \kappa;j+k+l-2m;-q).           
\end{align}

\begin{figure}[!t]
\includegraphics[width=1\linewidth]{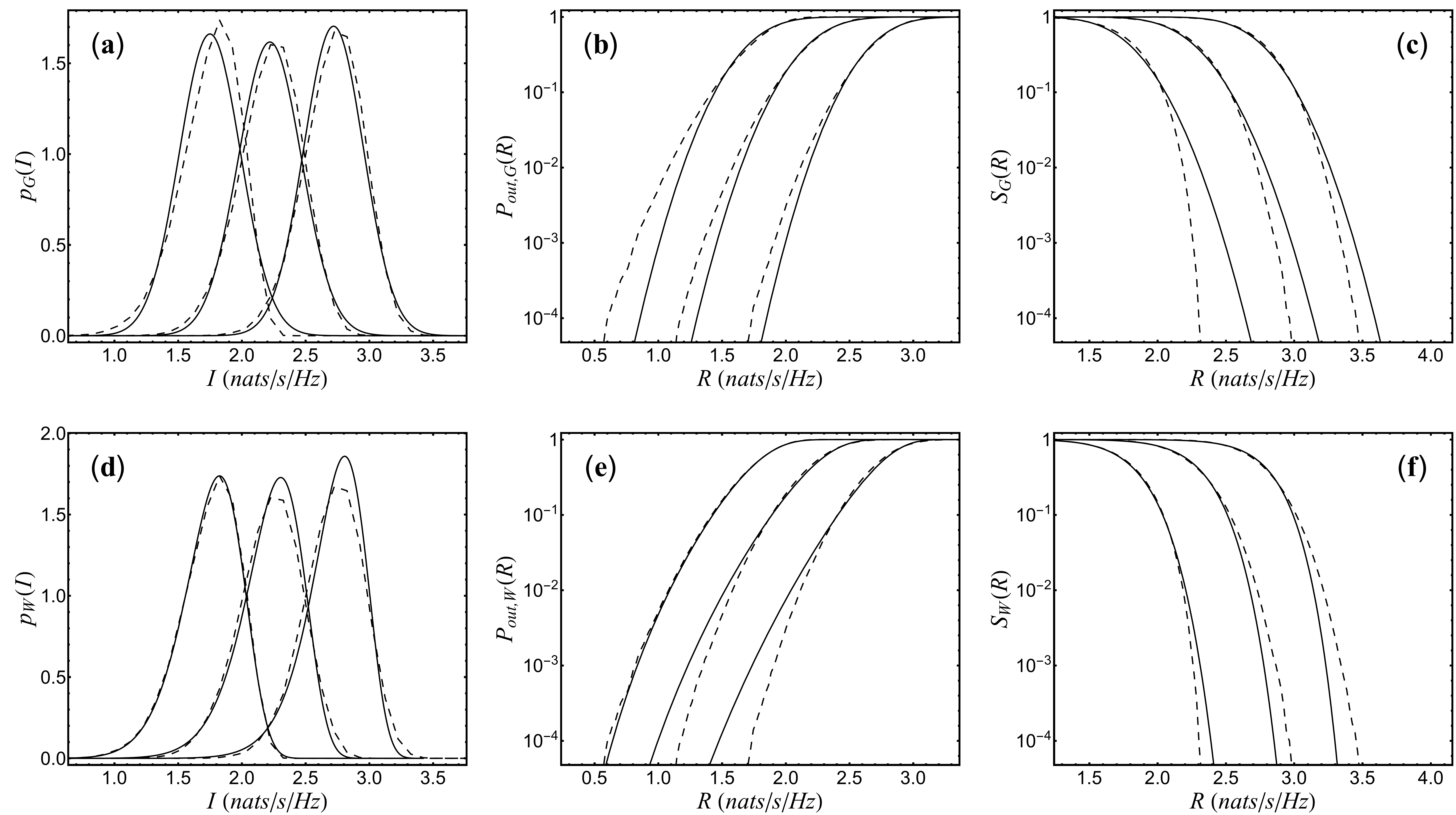}
\centering
\caption{\textit{Comparison of the Gaussian (a-c) and Weibull (d-f) approximations for PDF, CDF, and SF with Monte Carlo simulations by considering three sets of $(m, n, l)$ with $m\leq n$. In each plot, the parameters $(m, n, l)$ for the three curves, going from left to right are $(3, 6, 12)$, $(5, 7, 16)$ and $(7, 8, 17)$ respectively, and the $q_j$ values for $j=1,...m$, have been taken from the set $(q_1, q_2, q_3, q_4, q_5, q_6, q_7) = (8.80, 0.11, 0.09, 0.55, 1.20, 0.75, 0.28)$ depending on the value of $m$. The solid lines are based on the approximations and the dashed lines are from Monte Carlo simulations.}}
\label{fig4}
\end{figure}
\begin{figure}[!t]
\includegraphics[width=1\linewidth]{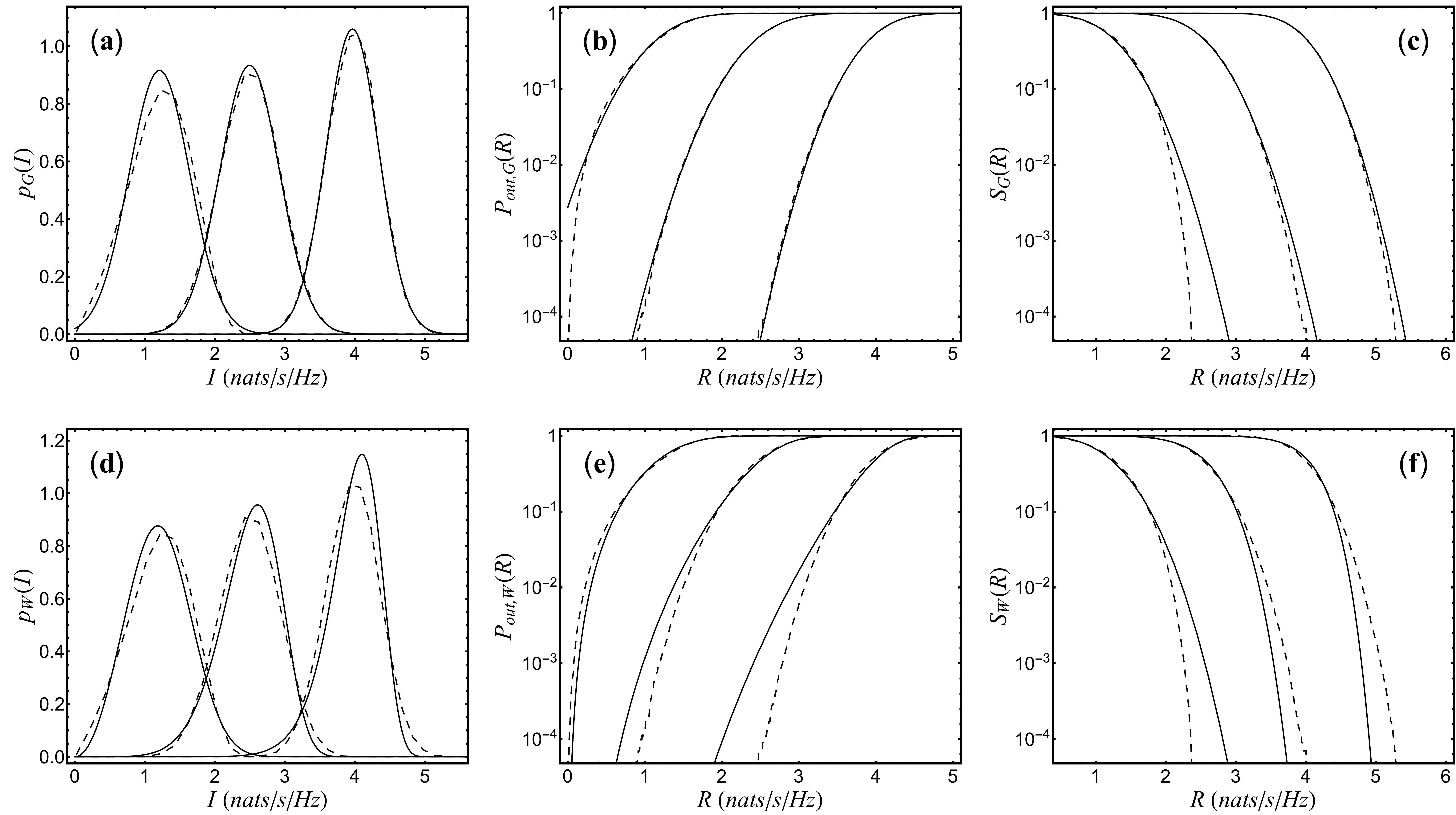}
\centering
\caption{\textit{Comparison of the Gaussian (a-c) and Weibull (d-f) approximations for PDF, CDF, and SF with Monte Carlo simulations by considering three sets of $(m, n, l)$ with $m>n$. In each plot, the parameters $(m, n, l)$ for the three curves, going from left to right are $(2, 1, 6)$, $(4, 3, 10)$ and $(7, 5, 14)$ respectively. The $q_j$ values for $j=1,...,m$ have been taken from the set $(q_1, q_2, q_3, q_4, q_5, q_6, q_7) = (11.00, 5.00, 1.50, 0.50, 0.75, 1.10, 3.30)$ depending on the value of $m$. The solid lines are based on the approximations and the dashed lines are from Monte Carlo simulations.}}
\label{fig5}
\end{figure}

%
\subsection{\textbf{Case-II- $ m>n $}}\label{sec2.2}
In this case, the matrix $ {\bf{H^\dag H}} $ possesses generic $n$ non-zero and  $(m-n)$ zero eigenvalues. We will handle the latter, encoded in the delta function in Eq.~\eqref{P2L}, in a limiting manner. We also note that each of the $m!$ permutations in Eq.~\eqref{P2L} lead to identical result when integrated over the eigenvalues. Therefore, we focus on one of those and multiply the resulting expression by $m!$. We obtain,
\begin{equation}
M(\kappa) = \lim_{\lambda_{n+1,...,\lambda_m}\to 0}\frac{C_{n,m}K_m(\kappa)}{\Delta_m(\{q\})}\int d\{\lambda\} \prod_{i=1}^n\lambda_i^{m-n} (1-\lambda_i)^{l-m-n}\Delta_{n}^2(\{\lambda\})\frac{\det[(1+q_j \lambda_k)^{\imath \kappa +m-1}]_{j,k=1}^m}{\Delta_m(\{\lambda\})}.
\end{equation}
It should be noted that the last factor in the above expression is of $0/0$ form and therefore we take the limit in the L'Hospital manner by operating $\prod_{j=n+1}^m (\partial ^{j-n-1}/\partial \lambda_j^{j-n-1})$ on both numerator and denominator and then setting ${\lambda_{n+1}=\cdots=\lambda_m=0}$~\cite{SMM2006,PKFD2017}. The differential operator with the variable $\lambda_j$ is acted upon the elements in the $j$th column. This gives us  
\begin{equation}
\label{Mklim}
M(\kappa) = \frac{C_{n,m}K_m(\kappa)}{\Delta_m(\{q\})G(m-n+1)}\int d\{\lambda\} \prod_{i=1}^n(1-\lambda_i)^{l-m-n} \Delta_{n}(\{\lambda\})\det[f_{j,k}(\kappa)]_{j,k=1}^m,
\end{equation}
where
\begin{align}
f_{j,k}(\kappa)=\begin{cases}
(1+q_j \lambda_k)^{\imath \kappa +m-1},~~~~~~~~~~~~~~~~~~~~~~~~~~~~~~~~~~ \text{for } k=1,...,n,\\
 (\imath \kappa +m+n-k+1)_{k-n-1} q_j^{k-n-1},~~~~~ \text{for } k=n+1,...,m,
\end{cases}
\end{align}
with $(a)_b=\Gamma(a+b)/\Gamma(a)$ being the Pochhammer symbol. Here, for the denominator we have used following result proved in the Appendix, 
\begin{equation}
\label{vandlim}
\prod_{j=n+1}^m\frac{\partial ^{j-n-1}}{\partial \lambda_j^{j-n-1}}\Delta_m(\{\lambda\})\Big|_{\lambda_{n+1}=\cdots=\lambda_m=0}=(-1)^{n(m-1)}G(m-n+1)\prod_{k=1}^n\lambda_k^{m-n}\,\Delta_n(\{\lambda\}),
\end{equation}
where $G(\eta)=\prod_{j=1}^{\eta-1} \Gamma(j) $ is the Barnes G-function.

The integrand in Eq.~\eqref{Mklim} contains a product over two determinants that involve $n$ and $m$ dimensional matrices, respectively. We use the standard expansion of the two determinants in terms of their respective elements and perform the integrals on the factors containing the $n (<m)$ eigenvalues. The resulting expression can be written as a determinant of an $m$-dimensional matrix. This is essentially a generalization of the Andr\'{e}ief's integral formula~\cite{Andreief1883,KG2010}. Afterwards, we absorb the $(-1)^{n(m-1)}$ factor arising in Eq.~\eqref{vandlim} within the final determinant by moving the first $n$ columns towards the end. We obtain eventually,
\begin{align}
\label{Mk2}
M(\kappa)  = \frac{n!\,C_{n,m}K_m(\kappa)}{\Delta_m(\{q\})G(m-n+1)}\det[h_{j,k}(\kappa)]_{j,k=1}^m,
\end{align}
where
\begin{align}
h_{j,k}(\kappa)=\begin{cases}
(\imath \kappa +m-k+1)_{k-1} q_j^{k-1},~~~~~~\text{ for }  k=1,...,m-n,\\
            B(k-m+n,l-m-n+1)~ {}_2F_1(k-m+n,1-m-\imath\kappa;k+l-2m+1; -q_j),\\
                  ~~~~~~~~~~~~~~~~~~~~~~~~~~~~~~~\text{ for } k=m-n+1,...,m.
   \end{cases}
\end{align}
The case when multiplicity occurs in $q_j$'s can be handled in the manner as outlined in the $m\le n$ case. Especially, when all the $q_j$'s are equal, the resultant MGF is again given by Eq.~\eqref{Mk1eq} but with $m$ and $n$ interchanged.

\section{Calculations of PDF and CDF}\label{sec3}

For the complete characterization of a random variable, one has to go beyond the moments or cumulants and analyze its PDF. Mutual information in the ergodic case being a random quantity, this holds for it also. The PDF is also required to evaluate the outage probability (CDF) which is a very useful metric for any communication channel. In fact, outage considerations are an integral part of space-division multiplexing system design in the optical communication~\cite{WF2011}.

In the following, based on moment or cumulant matching approach, we consider Gaussian and Weibull distribution models to obtain approximations for the PDF, CDF, and SF of the mutual information. The Gaussian distribution appears as a natural candidate when using the first two cumulants and has been shown to arise when a large number of channels is involved~\cite{MSS2003,HMT2004,HK2008}. Consequently, the Gaussian approximation for mutual information distribution has been used in earlier works as well and, in fact, has been found to work well even for not-so-large number of channels; see for example Refs.~\cite{PJ2002,PJ2008,ZG2004,PKF2017} in the context of MIMO wireless communication. On the other hand, our choice of Weibull distribution stems from the observation that the PDF of the mutual information for an arbitrary covariance matrix ${\bf Q}$ often exhibits negative skewness, as will be seen in the next section. Along with the above approximations, we also implement a very simple yet effective numerical Fourier inversion approach to obtain the PDF, CDF, and SF numerically from the MGF.

\begin{table}
\caption{KL divergence measures for comparison of Gaussian and Weibull approximations with the PDF calculated using Monte Carlo simulations. The $(m,n,l)$ values, with $m\le n$, correspond to the cases considered in Fig~\ref{fig4}.}
\begin{center}
  \begin{tabular}{|c|c|c|c|c|}
    \hline
    $m$ & $n$ & $l$ & $D_{KL}(p_{MC}||p_G)$ & $D_{KL}(p_{MC}||p_W)$ \\ \hline\hline
    $3$ & $6$ & $12 $ & $0.0392029$ & $0.00287793$\\ \hline
    $5$ & $7$ & $16 $ & $0.0107275$ & $0.0113861$ \\ \hline
    $7$ & $8$ & $17 $ & $0.00907784$ & $0.0228172$ \\
    \hline
  \end{tabular}
\end{center}
\label{table1}
\end{table}
\begin{table}
\caption{KL divergence measures for comparison of Gaussian and Weibull approximations with the PDF calculated using Monte Carlo simulations. The $(m,n,l)$ values, with $m> n$, correspond to the cases considered in Fig~\ref{fig5}.}
\begin{center}
  \begin{tabular}{|c|c|c|c|c|}
  \hline
   $m$ & $n$ & $l$ & $D_{KL}(p_{MC}||p_G)$ & $D_{KL}(p_{MC}||p_W)$ \\ \hline\hline
    $2$ & $1$ & $6 $ & $0.0165538$ & $0.0228172$ \\ \hline
    $4$ & $3$ & $10 $ & $0.00100549$ & $0.0185766$ \\ \hline
    $7$ & $5$ & $14 $ & $0.00112914$ & $0.0509409$ \\ \hline
  \end{tabular}
\end{center}
\label{table2}
\end{table}
\begin{table}
\caption{KL divergence measures comparing the numerical Fourier-inversion based PDF's with Monte Carlo simulation based PDF's for various choices of the cut-off length $L$ and discretization width (or step size) $\delta\kappa$. These results correspond to the plots in Figs.~\ref{fig6} and \ref{fig7}.}
\begin{center}
  \begin{tabular}{|c|c|c||c|c|c|}
      \hline
    \multicolumn{3}{|c||}{ $m\leq n$} &  \multicolumn{3}{|c|}{$m>n$}\\ \hline
    \hline
    $L$ & $\delta \kappa$ & $D_{KL}(p_{MC}||p)$ &     $L$ & $\delta \kappa$ & $D_{KL}(p_{MC}||p)$ \\ \hline\hline
    25 & 0.05 & $1.78572\times 10^{-4}$ &   15 & 0.05 & $4.18407\times 10^{-4}$  \\ \hline
    25 & 0.5 & $1.78697\times 10^{-4}$ &    15 & 0.5 & $4.18412\times 10^{-4}$ \\ \hline
    25 & 1 & $2.73943\times 10^{-4}$ &  15 & 1 & $4.18416\times 10^{-4}$ \\ \hline\hline
    22 & 0.05 & $1.32021\times 10^{-4}$ & 8 & 0.05 & $4.73338\times 10^{-4}$ \\ \hline
    28 & 0.05 & $2.05601\times 10^{-4}$ & 14 & 0.05 & $4.18386\times 10^{-4}$ \\ \hline
    34 & 0.05 & $2.29787\times 10^{-4}$ &   20 & 0.05 & $4.18411\times 10^{-4}$ \\ \hline
  \end{tabular}
\end{center}
\label{table3}
\end{table}

\subsection{Gaussian Approximation}\label{sec3.1}

Based on the Gaussian distribution, the PDF of the mutual information can be approximated using,
\begin{equation}
p_{G}(I)= \frac{1}{\sqrt{2\pi \sigma^2}}~ e^{-\frac{(I-\mu_1)^2}{2\sigma^2}},
\end{equation}
where $\mu_1$ is the mean and $\sigma^2 =\mu_2 - \mu_{1}^{2}$ is the variance of the MI which can be calculated by using Eq.~\eqref{mmnt}. We should remark that while the exact PDF of MI has support only in the non-negative domain $I\ge 0$, the above Gaussian approximation is nonzero even for $I<0$. However, the mean and variance of this Gaussian approximation are the ones corresponding to MI, which has the inherent support on the non-negative real axis. Therefore, evaluations of this approximation become negligible in the negative $I$ region, and consequently can be ignored. Additionally, the Gaussian approximation is not expected to capture the tail behavior of the exact PDF and holds good only within a finite number of standard deviations around the mean. Now, based on the above expression, the approximation for the outage probability or the CDF can be given by,
\begin{equation}
P_{out,G}(R) =\mathrm{Prob}(I<R) = \frac{1}{2}\left[1+ \text{erf}\left(\frac{R-\mu_1}{\sqrt{2\sigma^2}}\right)\right],
\end{equation}
where $\text{erf}(z)$ is the error function~\cite{AS1972}. Correspondingly, the survival function (SF) or reliability function for the Gaussian approximation is given by,
\begin{equation}
S_G(R)=\mathrm{Prob}(I>R) =\frac{1}{2}\left[1- \text{erf}\left(\frac{R-\mu_1}{\sqrt{2\sigma^2}}\right)\right].
\end{equation}
Evidently, we have $P_{out,G}(R)+S_G(R)=1$.

\begin{figure}[!t]
\includegraphics[width=0.9\linewidth]{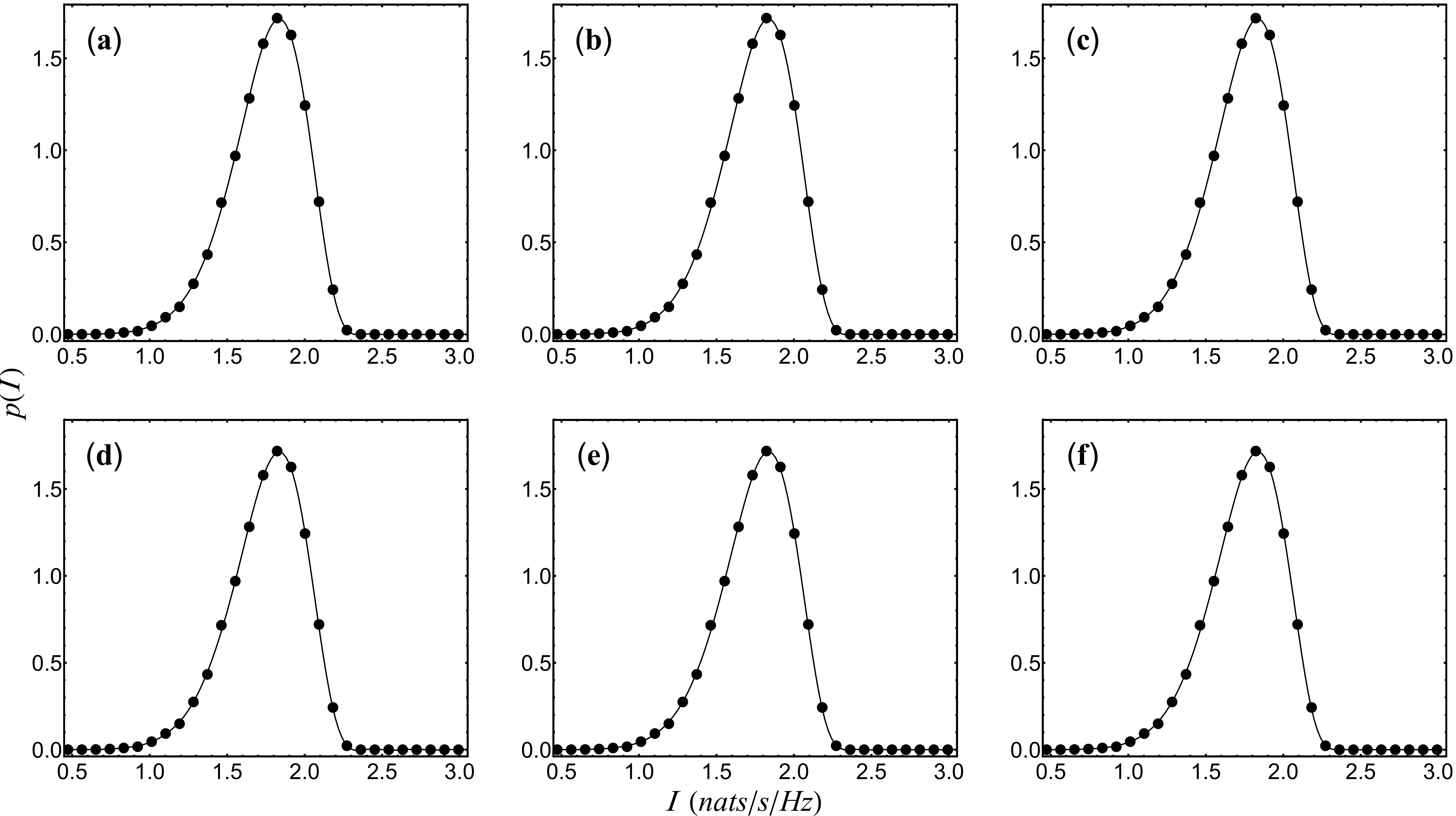}
\centering
\caption{\textit{Plots of PDF of MI calculated by numerical Fourier inversion technique for the case $m\le n$ with $(m,n,l)=(3,6,12)$, $(q_1,q_2,q_3) = (8.80, 0.11,0.09)$ and varying $L$ and $\delta\kappa$. In the first row, we have fixed $L=25$ and $\delta\kappa$ has been varied as (a) $\delta\kappa=0.05$, (b) $\delta\kappa=0.5$, (c) $\delta\kappa=1$. In the second row, $\delta\kappa=0.05$ is fixed and $L$ varies as (d) $L=22$, (e) $L=28$ (f) $L=34$. The solid lines are from analytical results and the disks are from Monte Carlo simulations. No noticeable difference is the observed between the plots, which is corroborated by extremely small KL divergence values as compiled in Table~\ref{table3}.}}
\label{fig6}
\end{figure}
\begin{figure}[!t]
\includegraphics[width=0.9\linewidth]{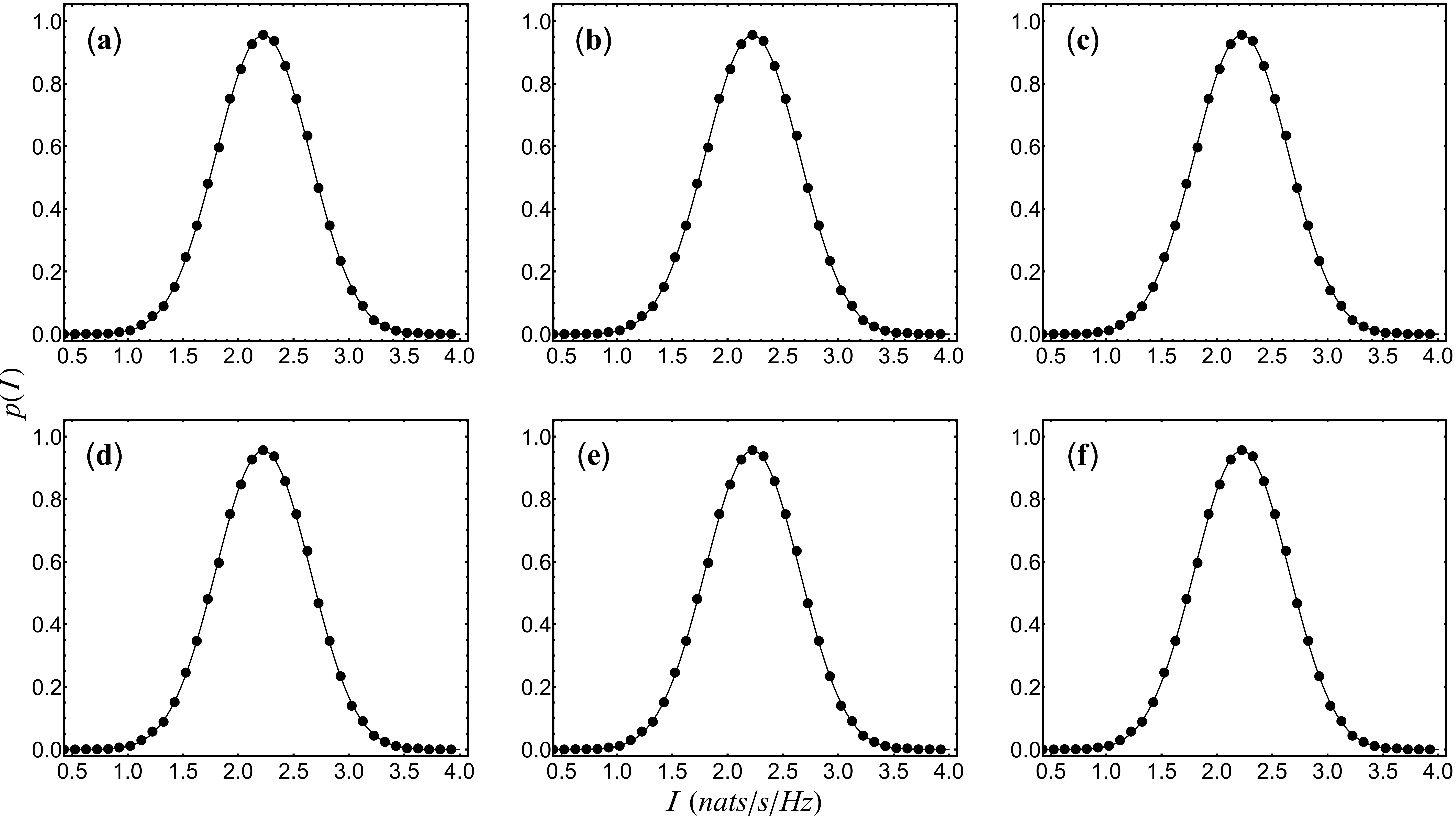}
\centering
\caption{\textit{
Plots of PDF of MI calculated by numerical Fourier inversion technique for the case $m> n$ with $(m,n,l)=(4,3,12)$, $(q_1,q_2,q_3,q_4) = (11.00, 5.00, 1.50, 0.50)$ and varying $L$ and $\delta\kappa$. In the first row, we have fixed $L=15$ and $\delta\kappa$ has been varied as (a) $\delta\kappa=0.05$, (b) $\delta\kappa=0.5$, (c) $\delta\kappa=1$. In the second row, $\delta\kappa=0.05$ is fixed and $L$ has been varied as (d) $L=8$, (e) $L=14$ (f) $L=20$. The solid lines are from analytical results and the disks are from Monte Carlo simulations. No noticeable difference is the observed between the plots, which is in conformity with the extremely small KL divergence values observed in Table~\ref{table3}.}}
\label{fig7}
\end{figure}

\subsection{Weibull Approximation}\label{sec3.2}

Using the Weibull distribution, we obtain the approximation for the mutual information PDF as
\begin{equation}
p_W(I)=
\frac{\beta}{\lambda}\left(\frac{I}{\lambda}\right)^{\beta-1}~e^{-\left(I/\lambda\right)^\beta},
\end{equation}
where the parameters $\beta$ and $\lambda$ are related to the first and second moments as,
\begin{equation}
\label{Wmoments}
\mu_1 = \lambda ~\Gamma\left(1+\frac{1}{\beta}\right), ~~
\mu_2 = \lambda^2 ~\Gamma\left(1+\frac{2}{\beta}\right).
\end{equation}
We note that the above PDF has its support on $[0,\infty)$, unlike the Gaussian approximation. We further note that in this case, the parameters $\beta$ and $\lambda$ need to be obtained numerically, or at least $\beta$ needs to be extracted by numerically solving $\mu_1^2/\mu_2=\Gamma^2(1+1/\beta)/\Gamma(1+2/\beta)$. The parameter $\lambda$ can then be obtained from either of the two relations in Eq.~\eqref{Wmoments}. Such calculations can be easily performed these days due to the availability of computational software packages, such as Mathematica and MATLAB. The outage probability or the CDF corresponding to the above approximation is obtained as
\begin{align}
P_{out,W}(R)=1-e^{-(R/\lambda)^\beta},
\end{align}
and the associated SF is given by,
\begin{equation}
S_W(R) =e^{-(R/\lambda)^\beta}.
\end{equation}

We will see in the next section that depending on the channel parameters, for some cases the Gaussian is more accurate and for some, the Weibull does better. For smaller number of channels, a crude strategy to decide which approximation may be better could be via the skewness, which can be evaluated using the moments. If the skewness is close to zero then Gaussian is a better choice and if it is significantly negative then very likely Weibull approximation would perform better. In the case of a large number of channels, the Gaussian approximation is expected to fare better~\cite{MSS2003,HMT2004,HK2008}.

\subsection{Numerical Fourier Inversion}\label{sec3.3}

From Eq.~\eqref{MGFMI}, it is clear that the exact PDF of the mutual information can be obtained by taking the inverse Fourier transform of the MGF. However, due to the complicated structure of the MGF, in general, it is not feasible to obtain an explicit expression for the PDF and therefore we resort to numerical Fourier inversion of the MGF. For this, we truncate the infinite domain by observing the behavior of the MGF using its numerical evaluation. An interval, say $[-L,L]$, is estimated in which it has a significant value. It is then discretized in terms of bins of width $\delta\kappa$, giving us the PDF as
\begin{equation}
\label{pIift}
p(I) = \frac{1}{2\pi}\int_{-\infty}^{\infty} e^{-\imath\kappa I}~ M(\kappa)~ d\kappa
       \approx \frac{1}{2\pi}\sum_{\kappa=-L}^{L} e^{-\imath\kappa I}~M(\kappa)~\delta\kappa.
\end{equation}
As examples, in Fig.~\ref{mgftrunc}, we illustrate two plots of MGF for the indicated parameter values and find that the MGF is negligible beyond $\pm 15$ and $\pm 10$, respectively. Therefore, accordingly, we may set the \emph{cut-off length} as $L=15$ and $L=10$ for numerical Fourier inversion in these two cases. In practice, one may decide the value of $L$ based on a  threshold, e.g. $M(\pm L)\approx 0.001$. As far as the choice of discretization width $\delta\kappa$ is concerned, for higher oscillations, smaller values of $\delta\kappa$ would be required for proper sampling. However, the use of very small $\delta\kappa$ can lead to numerical errors while performing the sum in Eq.~\eqref{pIift}. Remarkably, in our analyses we found that the final results are quite robust to the choices of both $L$ and $\delta\kappa$. This is discussed in the next section in more detail. The outage probability or the CDF can also be numerically obtained from the MGF as
\begin{equation}
\label{Pout}
P_{out}(R)= \mathrm{Prob} (I<R) = \int_{0}^{R} p(I)~dI\approx \frac{\imath}{2\pi} \sum_{\kappa=-L}^{L}\frac{( e^{-\imath\kappa R}-1)}{\kappa}~M(\kappa)~\delta \kappa,                 
\end{equation}
where $R$ is the required target rate. In the numerical evaluation of Eq.~\eqref{Pout}, $\delta\kappa$ should be chosen carefully so as to avoid the 0/0 occurence at $\kappa=0$. Alternatively, for $\kappa=0$, the factor $(e^{-\imath\kappa I}-1)/\kappa$ may be replaced by $-\imath I$. Similarly, the survival function or reliability function can be calculated as
\begin{equation}
S(R)=\mathrm{Prob}(I>R) \approx 1-\frac{\imath}{2\pi} \sum_{\kappa=-L}^{L}\frac{( e^{-\imath\kappa R}-1)}{\kappa}~M(\kappa)~\delta \kappa.
\end{equation}

\begin{figure}[!t]
\includegraphics[width=1\linewidth]{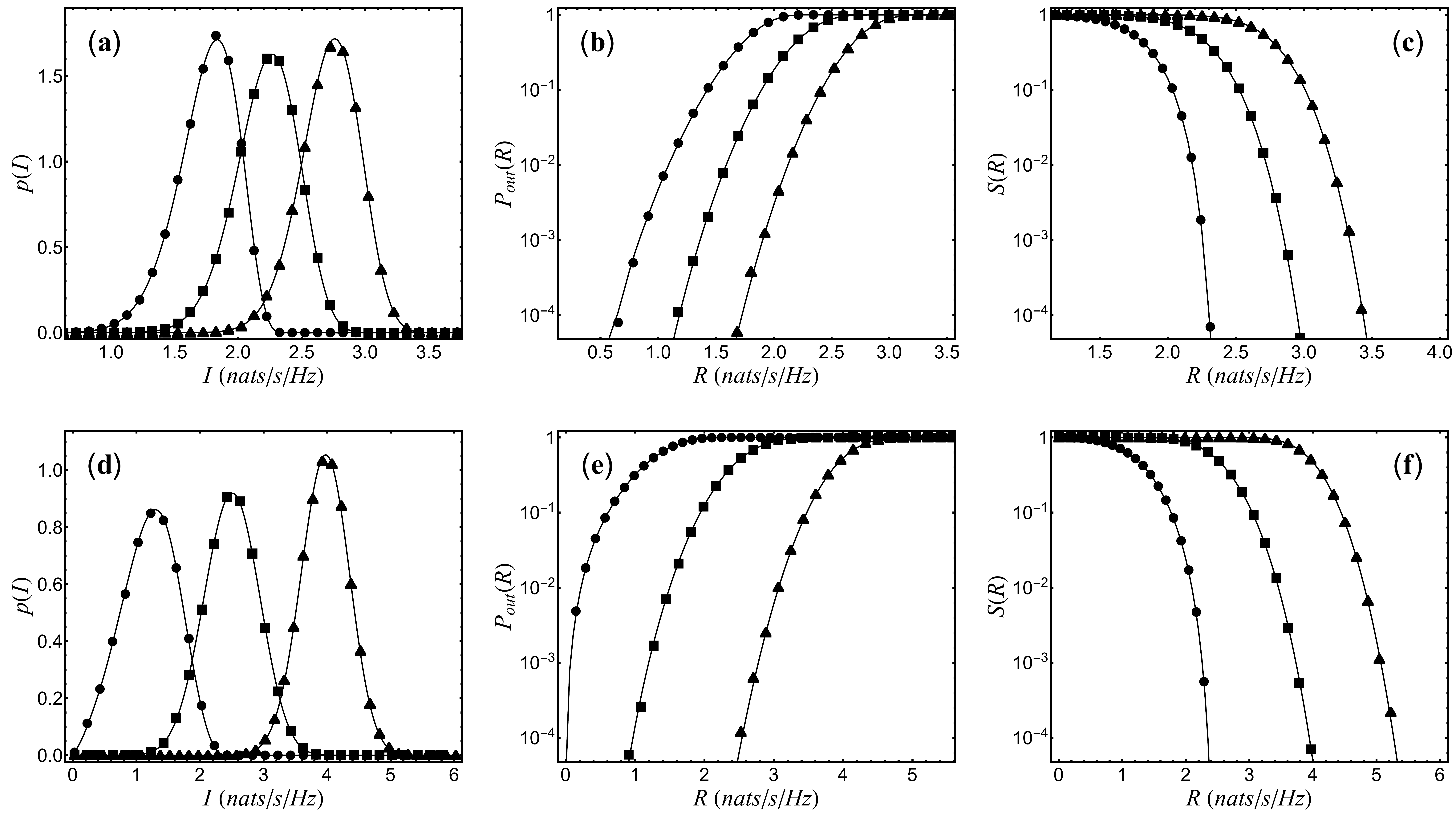}
\centering
\caption{\textit{Plots of the PDF, CDF, and SF based on numerical Fourier inversion, with varying $(m,n,l)$ values. In the first row, the case $m\leq n$ has been considered with $(m, n, l)$ for the three curves in each plot, going from left to right, taking values $(3, 6, 12)$, $(5, 7, 16)$ and $(7, 8, 17)$, respectively. The $q_j$ values $(j=1,...,m)$ are chosen from the set $(q_1, q_2, q_3, q_4, q_5, q_6, q_7) = (8.80, 0.11, 0.09, 0.55, 1.20, 0.75, 0.28)$ depending on the $m$ value. Similarly, in the second row the case $m>n$ has been considered. The parameters $(m, n, l)$ for the three curves in each plot in this row, going from left to right are $(2, 1, 6)$, $(4, 3, 10)$ and $(7, 5, 14)$, respectively. The $q_j$ values are taken from the set $(q_1, q_2, q_3, q_4, q_5, q_6, q_7) = (11.00, 5.00, 1.50, 0.50, 0.75, 1.10, 3.30)$ depending on $m$. The solid lines are based on the numerical Fourier inversion and for comparison, the symbols based on Monte Carlo simulations are also shown.}}
\label{fig8}
\end{figure}

\section{Results and Discussion}\label{sec4}

To begin with, in Figs.~\ref{approx1} and~\ref{approx2} we illustrate the behavior of the PDF, CDF (outage probability), and SF based on Gaussian (solid lines) and Weibull (dashed lines) approximations for both $m\le n$ and $m>n$ cases. The values of parameters used are indicated in the figure captions. Monte Carlo simulation results based on Eq.~\eqref{mutualinfo} are also shown as disks (filled circles.) We find that in Fig.~\ref{approx1}, the distribution is observed to be noticeably left (or negatively) skewed and the Weibull approximation agrees very well with the Monte Carlo simulation. The Gaussian approximation, having zero skewness, does not work that well in this case. In Fig.~\ref{approx2}, the skewness of the MGF probability density is close to zero, and the Gaussian approximation fares well compared to the Weibull approximation and is in close agreement with the Monte Carlo simulation. The skewness values calculated using the moments obtained from Eq.~\eqref{mmnt} for these two cases as $-0.62117$ and $-0.05385$, which conform to the observed behavior. To further quantify the difference of these approximate PDF's from the Monte Carlo simulation based probability density, we use the discretized version of the KL (Kullback--Leibler)-divergence measure~\cite{KL1951},
\begin{equation}
D_{KL}(p_{MC}||p)=\sum_i p_{MC}(I_i) \ln\left[\frac{p_{MC}(I_i)}{p(I_i)}\right]\delta I,
\end{equation}
Here, $p_{MC}(I)$ represents the PDF sampled using the Monte Carlo simulation, and $p(I)$ is the PDF to be compared with it. The smaller is the value of the KL divergence to 0, the better is the agreement between the compared PDF's. For our analyses, we implemented Monte Carlo simulation involving $2\times 10^5$ matrices to obtain the probability density of $I$. The points $I_i$ were sampled in the region where $P_{MC}(I)\gtrsim 10^{-2}$, Moreover, the separation between consecutive $I_i$ was taken as $\delta I=0.02$. For the parameters used in Fig.~\ref{approx1}, we obtained $D_{KL}(p_{MC}||p_G)=0.03867$ and $D_{KL}(p_{MC}||p_W)=0.00256$ which explains the better performance of Weibull approximation. For Fig.~\ref{approx2}, the evaluation gave $D_{KL}(p_{MC}||p_G)=0.00132$ and $D_{KL}(p_{MC}||p_W)=0.01789$ which in conformity with the observed better agreement of the Gaussian approximation with the empirical PDF.

\begin{figure}[!t]
\includegraphics[width=1\linewidth]{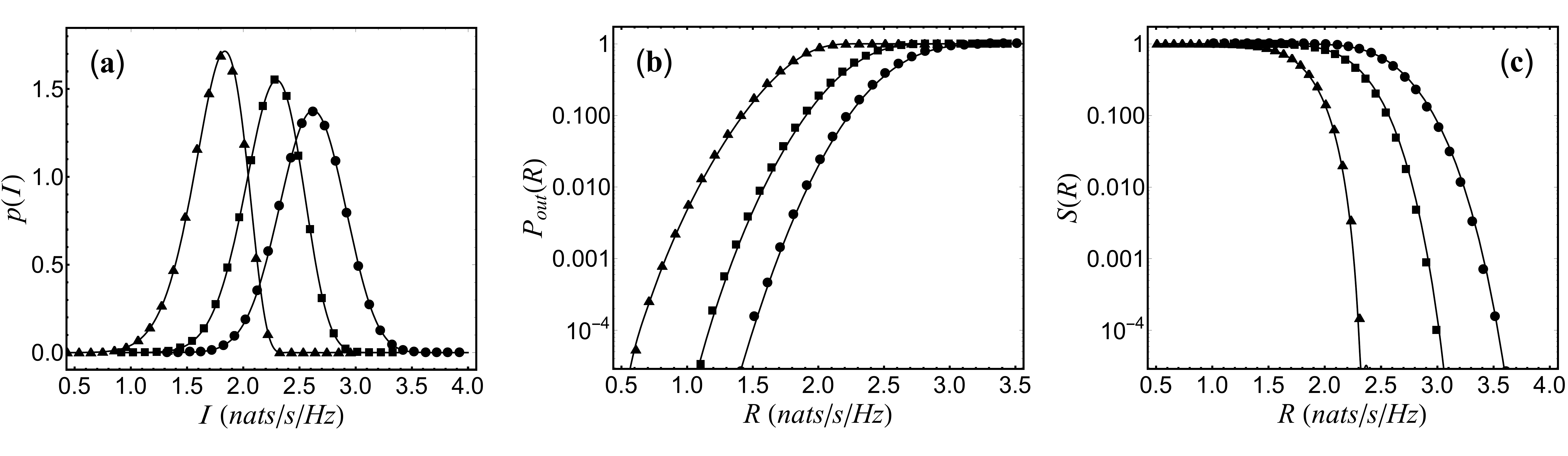}
\centering
\caption{\textit{Plots of PDF, CDF, and SF for the case $m\le n$, with $(m,n,l)=(3,6,12)$. The ($q_1,q_2,q_3$) values for the three curves, going from left to right, are $(8.80, 0.11, 0.09)$, $(6.80, 1.50, 0.70)$, and $(3, 3, 3)$, with fixed $\sum_jq_j=9$. The solid lines are using numerical Fourier inversion of the analytical MGF and the symbols are from Monte Carlo simulations.
}}
\label{fig9}
\end{figure}
\begin{figure}[!t]
\includegraphics[width=1\linewidth]{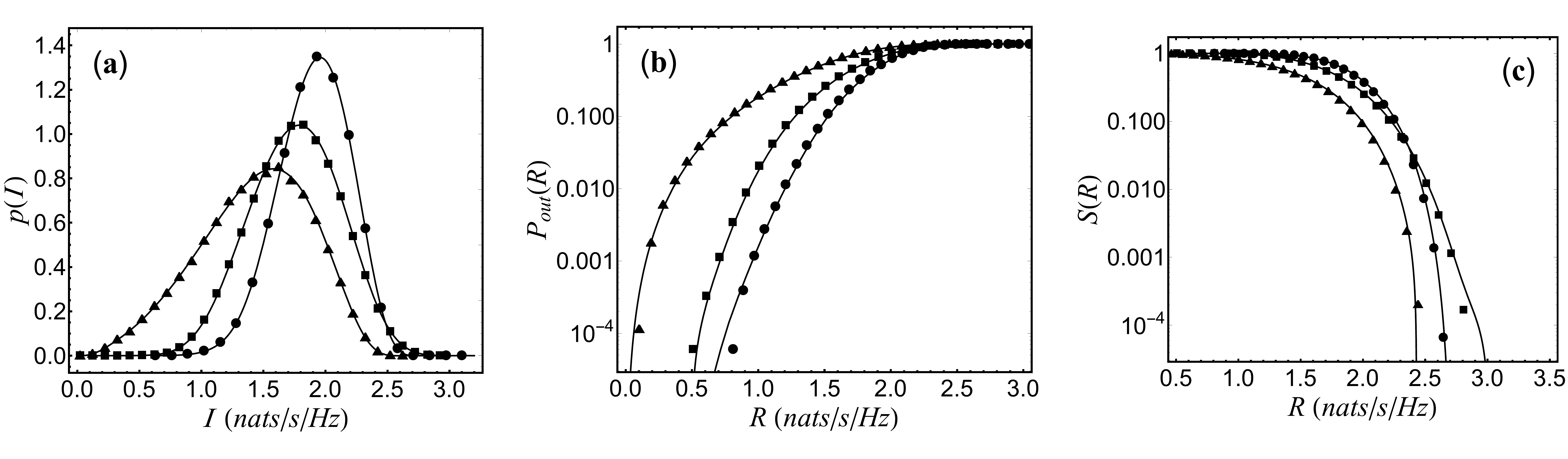}
\centering
\caption{\textit{Plots of PDF, CDF, and SF or the outage probability for the case $m>n$, with $(m,n,l)=(4,2,7)$. The ($q_1,q_2,q_3,q_4$) values for the three curves, going from left to right, are $(11.70,0.15,0.10,0.05)$, $(8.00, 2.00, 0.95, 1.05)$, and $(3, 3, 3, 3)$, with fixed $\sum_jq_j=12$. The solid lines are using numerical Fourier inversion of the analytical MGF and the symbols are from Monte Carlo simulations.}} 
\label{fig10}
\end{figure}

In Figs.~\ref{fig4} and~\ref{fig5}, we further explore the $m\le n$ and $m>n$ cases with varying $(m,n,l)$ values and choose $q_j$'s from a fixed set as indicated in the figure captions. For clarity, we examine the Gaussian and Weibull approximations (solid lines) separately in these plots and compare those with the Monte Carlo simulation results (dashed lines). We see that both the approximations work fairly well. The quantification of deviations in the bulk ($I$ values corresponding to $p_I(I)\gtrsim 10^{-2}$) using the KL divergence is compiled in Tables~\ref{table1} and \ref{table2}. The deviations in the tails can be observed in the outage probability and survival function plots.

We now consider our numerical Fourier-inversion approach and firstly examine the impact of variation in choices of the cut-off length $L$ and the discretization width $\delta\kappa$. In Fig.~\ref{fig6}, we consider the $m\le n$ case and plot the mutual information PDF's calculated using the Fourier-inversion approach for various combinations of $L$ and $\delta\kappa$. Similarly, in Fig.~\ref{fig7}, we examine the $m>n$ case for varying $L$ and $\delta\kappa$. For comparison, Monte Carlo simulation results are also shown. In both cases we find that no noticeable change in the respective PDF's are observed. The KL divergence values evaluated in Table~\ref{table3} for these different combinations indeed show tiny differences between the PDFs. Therefore, the robustness of the Fourier-inversion approach for evaluating the PDF of MI is implied and we use it in all the distribution related analyses that follow below.

In Fig.~\ref{fig8}, we examine the impact of variations of the parameters $l,m,n$ for both $m\le n$ and $m>n$ cases. In either case, the $q_j$ values are chosen from respective fixed sets, depending on the value of $m$ used. We observe perfect agreement between the Fourier-inversion based results and the Monte Carlo results for PDF, CDF as well as SF. Larger values of the channel dimension parameters indeed lead to increased mutual information and a higher threshold for the rate up to which the signal can be reliably communicated.

We now proceed to study the impact of the different choices of the transmission covariance matrix \textbf{Q} to study its effect on the mutual information statistics.
These results are shown in Figs.~\ref{fig9} and~\ref{fig10}. In these plots, the behavior of PDF, the CDF (outage probability), and the SF are exhibited for the indicated $m\le n$ and $m>n$ cases, respectively. As $\textbf{Q}$ is adjusted towards equal power condition, the expected feature of the MI curve shifting towards higher values can be clearly observed. Additionally, the skewness is seen to vary significantly making it farther or closer to a Gaussian-like behavior. The tail behavior of the distributions is highlighted in the log plots of CDF and SF. Monte Carlo simulation based results have been also displayed with the aid of various symbols, and they agree extremely well with our numerically Fourier-inverted results for the PDF, CDF, and SF. 

\begin{figure}[!t]
\includegraphics[width=0.9\linewidth]{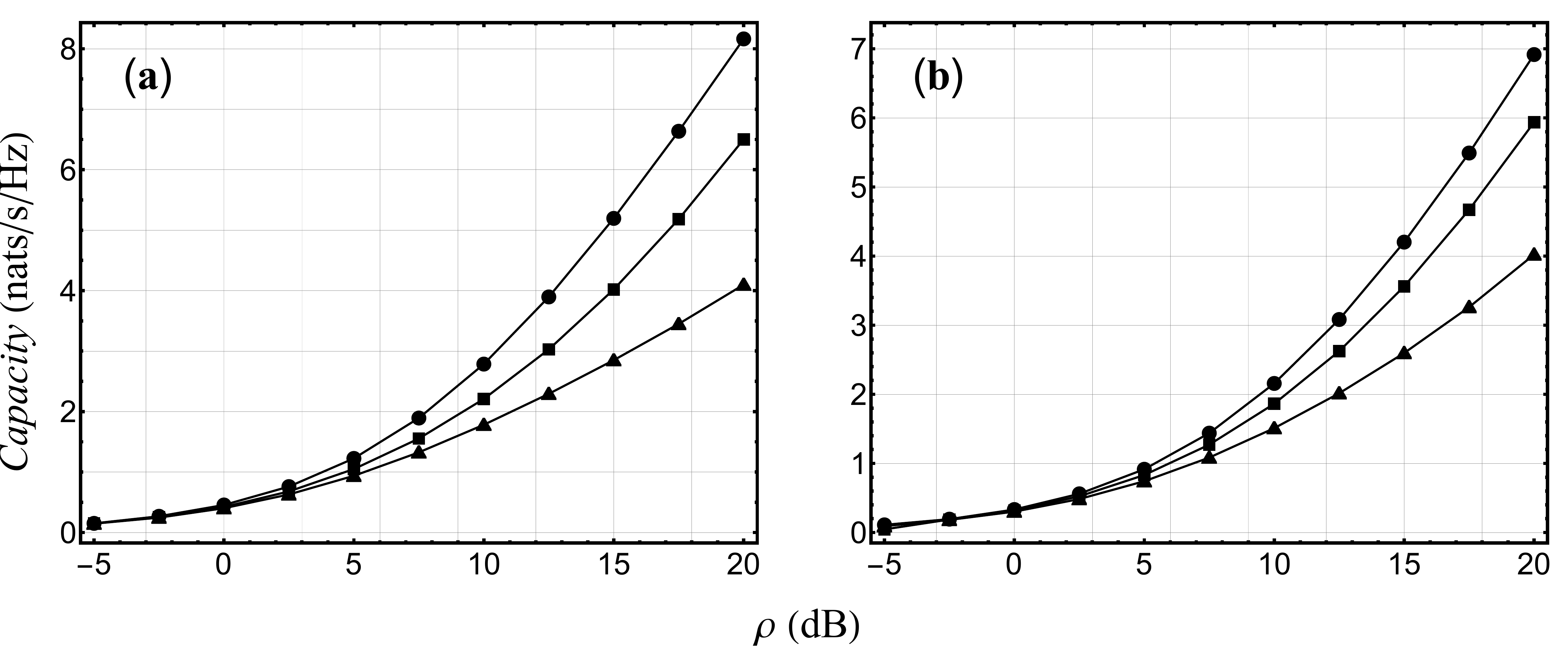}
\centering
\caption{\textit{Plots of ergodic capacity (mean value) of the mutual information: (a) $(m,n,l)=(3,6,12)$ and (b) $(m,n,l)=(4,3,10)$. For the three curves in (a), going from bottom to top, the ratio $(q_1:q_2:q_3)$ in absolute scale is $(0.995:0.003:0.002)$, $(0.850:0.100:0.050)$, and $(1:1:1)$. For the three curves in (b), going from bottom to top, the ratio $(q_1:q_2:q_3:q_4)$ is $(1.170:0.015:0.010:0.005)$, $(0.900:0.100:0.095:0.105)$, and $(1:1:1:1)$ in absolute scale. In both figures, the horizontal axis labels for $\rho=\sum_j q_j$ are depicted in decibel scale. The solid lines are using analytical result and the symbols are from Monte Carlo simulations.}}
\label{ergcap}
\end{figure}
\begin{figure}[!t]
\includegraphics[width=0.5\linewidth]{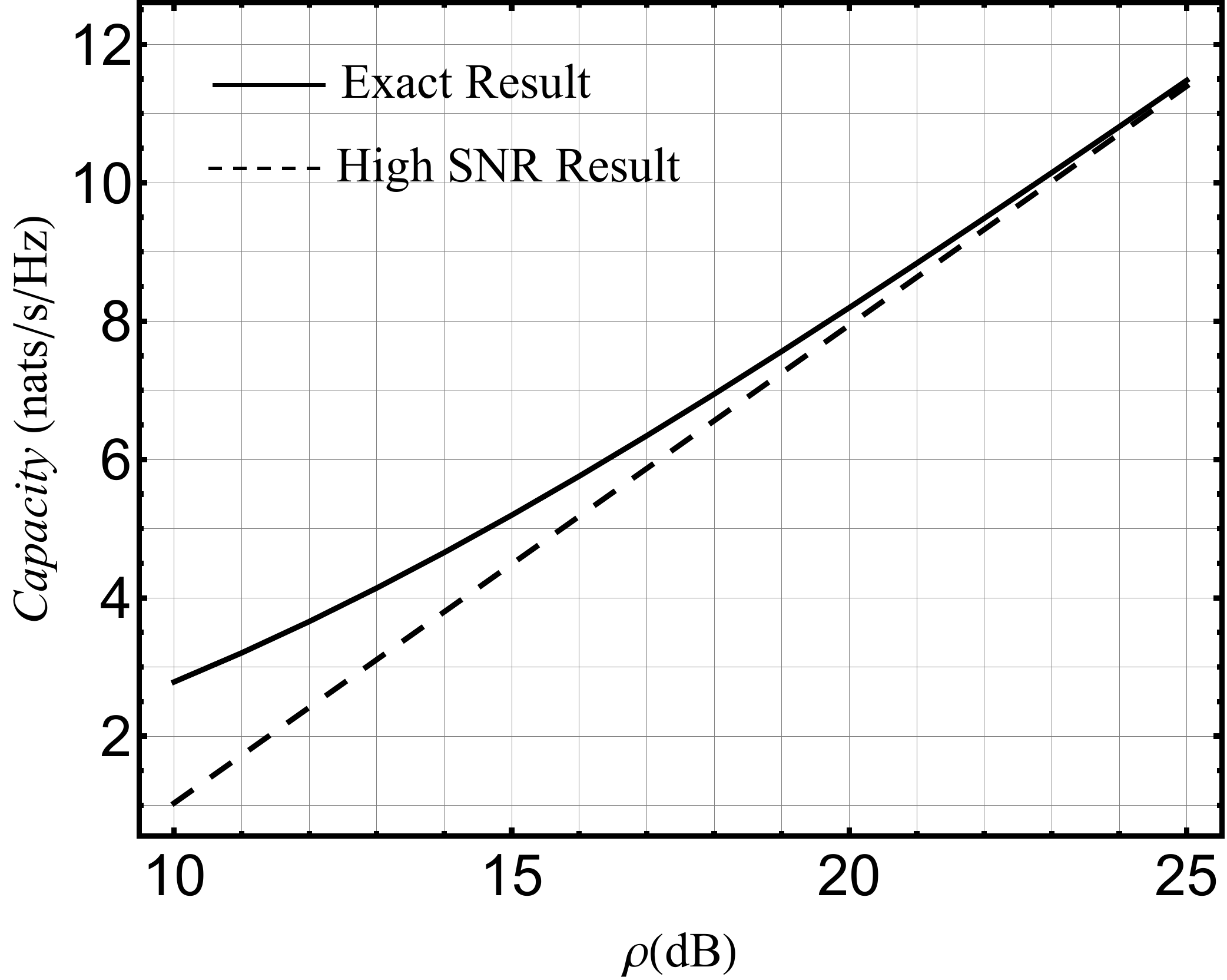}
\centering
\caption{\textit{Comparison of exact ergodic capacity result with the high SNR result given in Ref.~\cite{WLLB2019} for equal power case, i.e., ${\bf Q}=(\rho/m)\mathds{1}_m$. The parameters considered are $(m,n,l)=(3,6,12)$.}}
\label{comp}
\end{figure}

Next, we explore the behavior of the ergodic capacity. In Fig.~\ref{ergcap}, plots of ergodic capacity versus power are depicted using Eq.~\eqref{mmnt} with $\nu=1$ for various combinations of power allocation to distinct input channels, such that the total power $\rho=\text{tr}\,{\bf Q}=\sum_j q_j$ is fixed. Again, comparisons with Monte Carlo simulations based on Eq.~\eqref{mutualinfo} is also included, and we find excellent agreements. The equal power excitation case on each channel indeed leads to the highest value of the ergodic capacity. Finally, in Fig.~\ref{comp}, we consider an equal power allocation scenario and illustrate a comparison between our exact result for the ergodic capacity and the corresponding high-SNR result,
\begin{equation}
\mu_1\approx m \ln(\rho/m)+n \psi_0(n)-l\psi_0(l)-(n-m)\psi_0(n-m)+(l-m)\psi_0(l-m),
\end{equation} 
which has been recently derived in~\cite{WLLB2019}. Here, $\psi_0(z)$ is the digamma function~\cite{AS1972}. For the parameter values considered, we find that the latter is close to the exact result only for $\rho\gtrsim 25 \text{dB}$. Therefore, one should be careful in using the high SNR result.

\section{Conclusion}\label{sec5}

In this work, we have obtained an exact and closed-form result for the MGF of the mutual information for Jacobi-MIMO channel in the case of an arbitrary transmission covariance matrix ${\bf Q}$, which corresponds to unequal power associated with the excited modes used for signal transmission. The MGF has been further used to evaluate the ergodic channel capacity, PDF of the mutual information, CDF (outage probability), and SF with varying ${\bf Q}$ matrix. To this end, along with approximations based on Gaussian and Weibull distributions, we also proposed an effective numerical Fourier-inversion technique. All these results have been verified with Monte Carlo based numerical simulations. Our results help in understanding how unequal power allocation to the transmitting channels can impact the system performance. Therefore, we believe that our results constitute an important step in developing a better understanding of the under-addressed Jacobi MIMO channel.

\section*{Acknowledgments}

A.L. would like to thank Shiv Nadar University for providing financial support.

\section*{Appendix: Proof of Equation~\eqref{vandlim}}

Upon acting the derivatives on the $(n+1)$-th to $m$th column and then setting $\lambda_{n+1}=\cdots \lambda_m=0$, we obtain
\begin{equation*}
\prod_{j=n+1}^m\frac{\partial ^{j-n-1}}{\partial \lambda_j^{j-n-1}}\Delta_m(\{\lambda\})\Big|_{\lambda_{n+1}=\cdots=\lambda_m=0}=\det\begin{bmatrix}{\bf A} & {\bf B}\\{\bf C} & {\bf 0}\end{bmatrix}=(-1)^{n(m-1)}\det({\bf B})\det({\bf C}),
\end{equation*}
where ${\bf A}=[\lambda_{k}^{j-1}]_{j=1,...,m-n\atop k=1,...,n}$, ${\bf B}=[\Gamma(j)\delta_{j,k}]_{j,k=1,...,m-n}$, ${\bf C}=[\lambda_{k}^{j-1}]_{j=m-n+1,...,m\atop k=1,...,n}$, and ${\bf 0}=[0]_{m-n+1,...,m\atop k=n+1,...,m}$. Here, $\delta_{j,k}$ is the Kronecker delta function.
Now, it can readily be seen that $\det({\bf B})=\prod_{j=1}^{m-n}\Gamma(j)$, and $\det({\bf C})=\prod_{i=1}^n\lambda_i^{m-n}\det[\lambda_k^{j-1}]_{j,k=1,...,n}=\prod_{i=1}^n\lambda_i^{m-n}\,\Delta_n(\{\lambda\}),$ and therefore Eq.~\eqref{vandlim} follows.



\end{document}